\documentclass[a4paper,12pt]{article}
\usepackage[T2A]{fontenc}
\usepackage[cp1251]{inputenc}
\usepackage[english]{babel}
\inputencoding{cp1251}
\usepackage{indentfirst}
\usepackage[active]{srcltx}
\usepackage{graphicx}
\usepackage{float}
\usepackage[caption = false]{subfig}
\graphicspath{{./figure/}}
\usepackage[small,compact]{titlesec}
\usepackage{cite}
\usepackage{sfmath}
\usepackage{afterpage}
\usepackage{booktabs}
\usepackage{rotating}
\usepackage{multirow}

\oddsidemargin\evensidemargin
\newcommand{\D}{\displaystyle}
\newcommand{\acc}[1]{{\rm\accent'22\kern-0.5em}{#1}}
\newcommand{\ud}{\mathrm{d}}

\newcommand{\od}[3][]{\frac{\ud ^{#1}{#2}}{\ud {#3}^{#1}}}
\newcommand{\pd}[3]{\left(\frac{\partial{#1}}{\partial{#2}}\right)_{#3}}

\def\ni{\noindent}

\begin{document}
\title{Volume dependence of the Gr\"uneisen ratio for shock-wave equation-of-state studies}

\author{Valentin Gospodinov\\
Space Research and Technology Institute – BAS \\
BG-1000, Sofia, P.O. Box 799, BULGARIA\\
e-mail:  v.gospodinov@gmail.com}

\date{}
\maketitle

\begin{abstract}
This work presents an analysis of the existing self-contained expressions  for the volume dependence of the Gr\"uneisen ratio $\gamma$  in view of their further application to EOS (equation of state) studies. These expressions are assessed and applied to materials with the major types of chemical bonds. Predictions from regression analysis are compared to existing experimental data sets. All expressions predict with very good accuracy the values of $\gamma$ at ambient conditions, its volume variation in the low pressure region, but fail to give correct values for its infinite compression limit. A possible reason for this is that all experiments are performed at comparatively low pressures. The interpolation formula by Al'tshuler et al. (1987) and the equation, proposed by Jeanloz (1989) are the best fit to experimental data.  A modification to Jeanloz's equation, more convenient for use in shock physics, is proposed in the present work. It could be used jointly with the shock Hugoniot to derive a complete EOS for solids from their response to shock-wave loading.


\end{abstract}

\section*{Introduction}
\label{Intro}

The Gr\"uneisen ratio $\gamma$ is an important quantity in condensed
matter physics, shock physics and geophysics as it often occurs in
the research of thermodynamic behavior of matter at high pressures
and temperatures.

It may be used to estimate the value of the Debye characteristic
temperature ($\theta_D$) from the Debye-Gr\"uneisen definition of
$\gamma$ \cite[p.~133]{Girifalco_2000}

\[
d\ln \theta _D/d\ln V=-\gamma \quad \ \mbox{\rm or}\quad \ d\ln
\theta _D/dV=-\gamma /V.
\]
It is also important to predict the melting curve
\cite[p.~280]{Anderson_EoS}. Thermal EOS (equations of state)
require knowledge of the Gr\"uneisen ratio as well. In geophysics
the value of $\gamma$ is used to put constraints on geophysically
important parameters such as the pressure and temperature dependence
of the mantle and core, the adiabatic temperature gradient, and in the
geophysical interpretation of Hugoniot data \cite{Vocadlo_2000}.

The Gr\"uneisen ratio has both a statistical mechanics (microscopic)
and thermodynamic (macroscopic) definition. The former relates it to
the vibrational frequencies of the atoms in the crystal lattice of a
material, and the latter represents it in terms of well-known
thermodynamic properties. The experimental determination of $\gamma$
from its thermodynamic definition implies measurements of
thermodynamic properties at high pressures. The experimental
determination of the Gr\"uneisen ratio from its microscopic
definition is very difficult, since it requires a detailed knowledge
of the phonon dispersion spectrum of a material.

Because of the scarce experimental results and the lack of first
principle analytic equation, numerous phenomenological expressions
for the volume dependence of $\gamma$ have been reported in
literature. They predict a varying dependence of $\gamma$ as a
function of volume and some of them even give different values for
it at ambient pressure. Most of them are analyzed in two extensive
reviews --- by Knopoff and Shapiro \cite{Knopoff_rev}, and by
Anderson \cite{Anderson_rev}. Their accuracies are also compared in
recent works by X.\,Peng {\em et al}. \cite{Peng_2007} and by Cui
and Yu \cite{Guang-lei}. These papers are in the field of
geophysics. It is characteristic of them that there is an intrinsic
relationship between the expressions for $\gamma$, examined there,
and the cold or the normal isotherm. Many of these expressions
relate $\gamma$ at atmospheric pressure ($P=0$) to the first
derivative of the bulk modulus with respect to pressure or volume
($B_T'$).

To determine the functional dependence of the Gr\"uneisen ratio on
volume is a key problem in shock physics as well. Results from
shock-wave experiments provide direct information on the
compressional and thermal behavior of metals, ceramics, rocks, and
minerals at high pressures and high temperatures. Unfortunately,
data points are often sparsely deployed and irregularly distributed.
That is why it is a challenge, using this information, to have a go
on deriving the complete EOS for solids from their response to
shock-wave loading. Have it, one can easily obtain all their
thermodynamic properties by simple differentiation.

In this way, it is possible not only to obtain a reliable
interpolation tool, but to predict all compressional and thermal
properties of solids in the whole high-pressure high-temperature
region, attainable by shock-wave loading, standing on a sound
physical basis.

One of the ways to derive a complete EOS for solids from their
response to shock-wave loading is to use the specific form of this
dependence together with the shock Hugoniot. That is why it is
important to obtain the form of $\gamma$ independently of the shock
Hugoniot or of an isotherm.


To the author's knowledge, a comparison of the self-contained
expressions for the Gr\"uneisen ratio, used in shock physics, has
not been performed so far. Therefore, the objective of the present
work is to collect the most commonly used expressions for $\gamma$
and analyze and compare them to existing experimental data. It
differs from previous approaches \cite{Vocadlo_2000, Peng_2007,
Srivastava, Guang-lei} in that:
\begin{quote}
   $\circ $\ there is no intrinsic relationship between the expressions for $\gamma(V)$ analyzed here and the shock Hugoniot $P_H(V)$, the cold isotherm $P_c(V)$, or an arbitrary isotherm $P_T(V)$,

   $\circ $\ the expressions are applied to materials with various chemical bonds --- metallic ($Cu$, $\varepsilon$-$Fe$, $K$), ionic ($NaCl$), and covalent ($MgO$).
\end{quote}

\noindent The scope of the research with respect to the examined
materials and the maximum applied pressure is limited by the availability
of experimental data.

The paper is structured as follows. The question about the relation
between $\gamma(V)$ and $P_c(V)$ is clarified in
Sect.\,(\ref{volume_dependence}). In Sect.\,(\ref{shock_physics})
the most commonly used self-contained expressions for $\gamma(V)$,
mentioned above, are considered. In Sect.\,(\ref{LS_Fit}) regression
analysis of the experimental datasets is performed with these
expressions.The values of $\gamma$ at ambient conditions ---
\,$\gamma_0$, and at infinite pressure --- \,$\gamma_{\infty}$ are
treated as free parameters and are determined by the method of
least squares. The analysis and the discussion of the results
started in Sect.(\ref{LS_Fit}) is finalized in
Sect.(\ref{conclusions}). Also, conclusions are drawn there, and a
possible direction for continuing the research is outlined.

\section{The Gr\"uneisen ratio and the cold isotherm}
\label{volume_dependence}

The thermodynamic definition of the Gr\"uneisen ratio represents it
in terms of specific heat, thermal expansion coefficient, and bulk
modulus
\begin{equation}
\label{macro} \gamma = V\left(\frac{\partial P}{\partial E}\right)_V
= \frac{\alpha V B_T}{C_V} = \frac{\alpha V B_S}{C_P}\,,
\end{equation}
where $\alpha$ is the thermal expansion coefficient, $C_V$ -- the
specific heat at constant volume, $C_P$ -- the specific heat at
constant pressure, $B_T$ -- the isothermal bulk modulus, and $B_S$
-- the adiabatic bulk modulus. In terms of its thermodynamic
definition $\gamma$ may be considered  the measure of the change of
pressure resulting from the increase of internal energy at constant
volume. The experimental determination of $\gamma$, based on its
thermodynamic definition implies the concurrent measurement of the
involved thermodynamic properties at high pressures. 

The statistical mechanics definition relates it to the vibrational
frequencies of the atoms in the crystal lattice of a material
\begin{equation}
 \label{micro}
 \gamma _i=-\frac{V}{\nu _i}\left(\frac{\partial \nu _i}{\partial V}\right)_T = -\left(\frac{\partial \ln \nu _i}{\partial \ln V}\right)_T\quad (i=1,2,...,3N),
\end{equation}
where $\nu_i$ are the $3N$ vibrational frequencies of the crystal
lattice. The volume dependence of all lattice vibrational
frequencies is assumed one and the same
\cite[p.~130]{Girifalco_2000}, so

\begin{equation}
\label{Grun_assumpt} \gamma = -\left(\frac{\partial \ln
\nu}{\partial \ln V}\right)_T\,.
\end{equation}

In principle, if we knew the interatomic potential, we could
calculate the frequency spectrum of the crystal and its change with
volume and in this way specify the form of $\gamma(V)$. However,
this problem is mathematically so complex that has not been solved
so far. That is why we have to resort to various model concepts to
obtain the volume dependence of $\gamma$.

{\sl{Slater – Landau formula}}.
 Slater\,\cite{Slater_1939bk} and Landau \cite{Landau_1945} derived the following expression for the Gr\"uneisen ratio on the basis of a model of an elastic medium, and assuming that the Poisson ratio does not vary with volume:

    \begin{equation}
\gamma = -\frac{V}{2}\,\frac{\ud^2 P_c/\ud V^2}{\ud P_c/\ud V} -
\frac{2}{3}\,,
    \label{Slater_Landau}
    \end{equation}
where $P_c=P_c(V)$ is the cold compression curve.

Later Slater\,\cite{Slater_1940} and Gilvarry\,\cite{Gilvarry_1956},
using the values of the first and second derivatives at zero
pressure (derived from Bridgman’s data on compressibility)
calculated $\gamma$ from Eq.\,(\ref{Slater_Landau}) and compared it
to thermodynamic values of the Gr\"uneisen ratio. A good agreement
was obtained for the majority of metals.

{\sl{The Dugdale – MacDonald relation}}. Dugdale and MacDonald in a
short note\,\cite{Dugdale_1953} proposed to modify
Eq.(\ref{Slater_Landau}) and wrote the following expression for the
Gr\"uneisen parameter:

    \begin{equation}
\gamma = -\frac{V}{2}\,\frac{\ud^2\left(P_c V^{2/3} \right)/\ud
V^2}{\ud\left(P_c V^{2/3}\right)/\ud V} -\frac{1}{3}.
    \label{Dugdale_MacDonald}
    \end{equation}
However, there were some erroneous assumptions in their reasoning.
Subsequently, Rice {\sl et al}\,\cite{Rice_1958} have proposed a
somewhat different derivation for Eq.\,(\ref{Dugdale_MacDonald}).
They obtained it for a cubic lattice, assuming that all force
constants depend on volume in the same way. The values of $\gamma$,
obtained from Eq.\,(\ref{Dugdale_MacDonald}) at zero pressure, are
in good agreement with the thermodynamic values of $\gamma_0$.

{\sl{The Gr\"uneisen ratio in the free volume approximation}}.
Zubarev and Vashchenko\,\cite{Vaschenko_1963} studied the vibration
of atoms in the spherically symmetric field of their neighbors (the
free volume theory). They have obtained the following expression for
the Gr\"uneisen parameter:
    \begin{equation}
\gamma = -\frac{V}{2}\,\frac{\ud^2\left(P_c V^{4/3} \right)/\ud
V^2}{\ud\left(P_c V^{4/3}\right)/\ud V}.
    \label{Free_volume}
    \end{equation}

{\sl{Generalized formula for $\gamma$}}. All three equations
(\ref{Slater_Landau}), (\ref{Dugdale_MacDonald}) and
(\ref{Free_volume}) can be combined into one

    \begin{equation}
\gamma = -\frac{V}{2}\,\frac{\ud^2\left(P_c V^{2m/3} \right)/\ud
V^2}{\ud\left(P_c V^{2m/3}\right)/\ud V} +\frac{m-2}{3},
    \label{combined}
    \end{equation}

\ni which at m = 0 turns into Eq.\,(\ref{Slater_Landau}), for m = 1
--- into Eq.\,(\ref{Dugdale_MacDonald}), and for m = 2 --- into
Eq.\,(\ref{Free_volume}). At atmospheric pressure (taken as $P=0$)
all expressions for the Gr\"uneisen ratio depend solely on
$B'_{\scriptscriptstyle T_0}$ ($\ud B_{\scriptscriptstyle T}/\ud P$
at $P=0$).
\[
\D \gamma_{\scriptscriptstyle S}=\frac{1}{2}B'_{\scriptscriptstyle
T_0}-\frac{1}{6}, \,\gamma_{\scriptscriptstyle
{DM}}=\frac{1}{2}B'_{\scriptscriptstyle T_0}-\frac{1}{2},
\,\gamma_{\scriptscriptstyle {VZ}}=\frac{1}{2}B'_{\scriptscriptstyle
T_0}-\frac{5}{6}.
\]
They give different values for the Gr\"uneisen ratio which are
connected to each other by the following relation:

    \begin{displaymath}
\gamma_{m=0} = \gamma_{m=1} + \frac{1}{3} = \gamma_{m=2} +
\frac{2}{3}.
    \end{displaymath}

It can be readily seen from
Eqs.\,(\ref{Slater_Landau})\,-\,(\ref{combined}) that within the
framework of these approaches the Gr\"uneisen ratio is a direct
function of the chosen cold isotherm, which defines $P_c$ as a
function of $V$.

Expressions for $\gamma (V)$, independent of the cold isotherm, the
normal isotherm, and the shock Hugoniot are considered in the next
section.
\section{The Gr\"uneisen ratio in shock physics}
\label{shock_physics}

There are a plethora of stand-alone expressions for the Gr\"uneisen
ratio which predict a varying dependence of $\gamma$ on volume.
Perhaps the most commonly used one is the empirical law
\begin{equation}
\label{by_Anderson} \gamma\rho^q = const,
\end{equation}
\noindent proposed by Anderson \cite{Anderson_1979}, where $q$ is a
quantity near unity. In his paper \cite{Anderson_1979} Anderson
considers the possible application of Eq.(\ref{by_Anderson}) for
temperature calculations pertaining to the lower mantle. He points
out that a value of $q$ anywhere in the range $0.8<q<2.2$ is
acceptable on the basis of seismic data. Using regression analysis,
Anderson shows that the '$\gamma\rho^q = const$' approximation is
not sensitive to the choice of $q$ if it is in the above interval.
In his opinion this is due to the restricted range of compression,
corresponding to the lower mantle. He chooses the simplest function
(with $q=1$)
\begin{equation}
\label{by_Anderson_1} \gamma\rho = const,
\end{equation}
\noindent and further assumes that
\begin{equation}
\label{common} \gamma\rho = \gamma_0\rho_0.
\end{equation}
\noindent This expression seems to be the most commonly used in
shock physics. As to its validity it is usually stated that
Eq.(\ref{common}) is valid for not very high pressures. Actually,
Anderson \cite{Anderson_1979} lays stress on the narrow range of
compression, corresponding to the lower mantle, not on the absolute
value of the specific pressure.

Many authors \cite{Bennet_1978, Thomson_1972, Royce_1971} have
combined Eq.(\ref{common}) with the fact that at large compressions
the limiting value of $\gamma$ for all materials is that of the
degenerate electron gas\,\cite{Holzapfel_2001}, namely ($2\over3$),
to write down interpolation formulae for the volume variation of
$\gamma$. Some of these are
\begin{equation}
\label{LANL} \gamma=\gamma_0\rho_0/\rho +
\frac{2}{3}(1-\rho_0/\rho),
\end{equation}

\begin{equation}
\label{Sandia}
\gamma=\gamma_0\rho_0/\rho+\frac{2}{3}(1-\rho_0/\rho)^2,
\end{equation}

\begin{equation}
\label{LLNL} \gamma=\gamma_0 - a(1-\rho_0/\rho).
\end{equation}

Al'tshuler et al.\,\cite{25_metals} have proposed the following
expression:
\begin{equation}
\label{Altshuler} \gamma=\gamma_{\infty} + (\gamma_0 -
\gamma_{\infty})/\sigma^m,
\end{equation}
where $\gamma_{\infty}$ = $2\over3$ for all elements except alkali
elements, for which $\gamma_{\infty}$ = $1\over2$.  In
Eqs.(\ref{LANL})\,-\,(\ref{Altshuler}) $a$ is a material dependent
constant, $\sigma = \rho/\rho_0$, $m = \gamma_0/(\gamma_0 -
\gamma_{\infty})$, and $\gamma_0$ and $\gamma_{\infty}$ are the
values of $\gamma$ at ambient conditions and at infinite pressure,
respectively. According to the authors of\,\cite{25_metals} the
logarithmic derivative of Eq.(\ref{Altshuler}) is close to the
experimental derivative and the asymptotic values $\gamma_{\infty}$
correspond to the quantum-statistical Gr\"uneisen coefficients of
the crystal lattice under extreme degrees of
compression\,\cite{Kopyshev_1965}.

Jeanloz \cite{Jeanloz_1989}, starting from the second Gr\"uneisen
ratio,
\begin{equation}
\label{Gruneisen_2} q = \pd{\ln\gamma}{\ln V}{T},
\end{equation}
assumed it to depend on volume only. The particular volume
dependence he used is given by
\begin{equation}
\label{q_Jeanloz} q = q_0 \left(\frac{V}{V_0}\right)^{q'}.
\end{equation}
The logarithmic derivative of $q$,
\begin{equation}
\label{q_prim} q' = \od{\ln q}{\ln V},
\end{equation}
known as the third Gr\"uneisen ratio, is supposed to be a
material-dependent constant.

Then, for the particular volume dependence of $\gamma$, Jeanloz
obtained
\begin{equation}
\label{Jeanloz_g} \gamma = \gamma_0
exp\left\{\left(\frac{q_0}{q'}\right)\left[\left(\frac{V}{V_0}\right)^{q'}
- 1\right]\right\},
\end{equation}
where $\gamma_0$, $q_0$, and $V_0$ are the values of $\gamma$, $q$,
and $V$ at ambient conditions.

Srivastava and Sinha \cite{Srivastava} modify Eq.(\ref{Jeanloz_g})
to introduce in it the infinite compression limit of $\gamma$. They
assume $\gamma_{\infty}$=($1\over2$). For $P \to \infty$, i.e. $V
\to 0$, Eq.(\ref{Jeanloz_g}) yields
\begin{equation}
\label{g_infty} \gamma_{\infty} = \gamma_0
exp\left(-\frac{q_0}{q'}\right).
\end{equation}

Now, following the model of an oscillating lattice of ions in a
uniform neutralizing background of electrons, Eq.(\ref{g_infty})
gives
\[
\label{g_inf_half} \gamma_0 exp\left(-\frac{q_0}{q'}\right) =
\frac{1}{2},
\]
or $q_0/q' = \ln(2\gamma_0)$. Then, Eq.(\ref{Jeanloz_g}) takes the
form
\begin{equation}
\label{Jeanloz_g1} \gamma = \gamma_0
exp\left\{\ln(2\gamma_0)\left[\left(\frac{V}{V_0}\right)^{q'} -
1\right]\right\}.
\end{equation}
This equation satisfies the infinite compression limit for $\gamma$,
i.e. at $P\to\infty$ or $V\to0$, $\gamma = \gamma_{\infty}$ =
($1\over2$).

Other researchers \cite{Bennet_1978, Thomson_1972, Royce_1971,
Holzapfel_2001} have favored for solids
$\gamma_{\infty}$=($2\over3$) which follows from the degenerate
electron gas model. Therefore, Eq.(\ref{g_infty}) with
$\gamma_{\infty}$ = ($2\over3$) should be considered as well. With
($2\over3$) as the infinite compression limit in
Eq.(\ref{Jeanloz_g}), we have
\begin{equation}
\label{Jeanloz_g2} \gamma = \gamma_0
exp\left\{\ln\left(\frac{3}{2}\gamma_0\right)\left[\left(\frac{V}{V_0}\right)^{q'}
- 1\right]\right\}.
\end{equation}

Here I propose a general form of Eq.(\ref{Jeanloz_g}) which
incorporates both Eqs.(\ref{Jeanloz_g1}) and (\ref{Jeanloz_g2})
\begin{equation}
\label{my_gamma} \gamma = \gamma_0
exp\left\{\ln\left(\frac{\gamma_0}{\gamma_{\infty}}\right)\left[\left(\frac{V}{V_0}\right)^{q'}
- 1\right]\right\}.
\end{equation}
In this equation $\gamma_0$, $\gamma_{\infty}$, and $q'$ are treated
as free parameters and will be determined by regression analysis of
the experimental data sets.

Rice has also derived an expression for $\gamma$ \cite{Rice_1965}
based on its thermodynamic definition. He makes two assumptions:
first, that the Gr\"uneisen ratio $\gamma=V(\partial P/\partial
E)_V$is a function of volume only; and second, that the adiabatic
bulk modulus $B_S=-V((\partial P/\partial V)_S$ is also a function of
volume only. His expression has the form:
\begin{equation}
\label{Rice_1} (V_0/V)\gamma = (\varepsilon + 1/\gamma_0)^{-1}.
\end{equation}
After some rearrangements we obtain:
\begin{equation}
\label{Rice_2} \gamma = \gamma_0(1 - \varepsilon)(1 +
\gamma_0\varepsilon)^{-1},
\end{equation}
where $\varepsilon = 1 - V/V_0$ is the dimensionless volume.

Equations\,(\ref{Rice_1}) and (\ref{Rice_2}) give incorrect value
for $\gamma_{\infty}$, i.e. $'0'$ and fail to describe adequately
any of the datasets used here. That is why they are excluded from
further consideration. The results from the calculations and a
comparison of the other expressions are presented in the next
section. The values of $\gamma_{\infty}$, obtained by regression
analysis, are given careful consideration there as well.
\section{Fitting the expressions for $\gamma(V)$ to experimental data}
\label{LS_Fit}

The experimental points for the regression analysis of the models
(Eqs.(\ref{common})\,-\,(\ref{Altshuler}), (\ref{Jeanloz_g}),
(\ref{Jeanloz_g1})\,-\,(\ref{my_gamma})) are taken from
\cite{25_metals, Birch_1986, Anderson_2001, Anderson_2001c,
Boehler_1983, Anderson_1993}. In these papers diverse variables are
used for the volume dependence of $\gamma$ --- $\rho/\rho_0$,
$\eta=V/V_0$, $\varepsilon=1 - V/V_0$. In the present work the
relative volume $\varepsilon=1 - V/V_0$ is introduced in all models.
The original and the transformed expressions are given in
Tabl.\,(\ref{all_equations}). \afterpage{\clearpage}
\begin{sidewaystable}
\centering
\begin{tabular}{@{}p{2.5cm}p{7cm}p{7cm}p{1cm}@{}}
\toprule[1pt] \multicolumn{1}{l}{\multirow{2}{*}{Reference}} &
\multicolumn{2}{c}{Expressions for $\gamma$}\\
\cmidrule[0.6pt](lr){2-3}& Original equations & In this work
&\multicolumn{1}{r}{Eq.}\\
\midrule[1pt]

Anderson\,\cite{Anderson_1979} & $\gamma\rho=\gamma_0\rho_0=const$ & $\gamma=\gamma_0(1-\varepsilon)$ & (\ref{common})\\[1ex] \\ [-1.5ex]

Bennett\,{\em et al}\,\cite{Bennet_1978} & $\gamma=\gamma_0\rho_0 /\rho+(2/3)(1-\rho_0 /\rho)$ & $\gamma=\gamma_0(1-\varepsilon)+\gamma_{\infty}\varepsilon$ & (\ref{LANL})\\[1ex] \\ [-1.5ex]

Thomson\,\,and Lauson\,\cite{Thomson_1972} & $\gamma=\gamma_0\rho_0/\rho+(2/3)(1-\rho_0/\rho)^2$ & $\gamma=\gamma_0(1-\varepsilon)+\gamma_{\infty}\varepsilon^2$ & (\ref{Sandia})\\[1ex] \\ [-1.5ex]

Royce\,\cite{Royce_1971} & $\gamma=\gamma_0 - a(1-\rho_0/\rho)$ & $\gamma=\gamma_0-a\varepsilon$ & (\ref{LLNL})\\[1ex] \\ [-1.5ex]

Al'tshuler {\em et~al}\,\cite{25_metals} & $\gamma=\gamma_{\infty} +
(\gamma_0 - \gamma_{\infty})/\sigma^m,\,
    \sigma = \rho/\rho_0,\,m = \gamma_0/(\gamma_0 - \gamma_{\infty})$ & $\gamma=\gamma_{\infty}+(\gamma_0-\gamma_{\infty})(1-\varepsilon)^m,\,
    m = \gamma_0/(\gamma_0-\gamma_{\infty})$ & (\ref{Altshuler})\\[0.8ex] \\ [-1.5ex]
Jeanloz\,\cite{Jeanloz_1989} & $\gamma = \gamma_0
exp\left\{\left(\frac{q_0}{q'}\right)\left[\left(\frac{V}{V_0}\right)^{q'}
- 1\right]\right\}$ & $\gamma = \gamma_0
exp\left\{\left(\frac{q_0}{q'}\right)\left[\left(1-\varepsilon\right)^{q'}
- 1\right]\right\}$ & (\ref{Jeanloz_g})\\[0.8ex] \\ [-1.5ex]
Srivastava\,\,and Sinha\,\cite{Srivastava} & $\gamma = \gamma_0
exp\left\{\ln(2\gamma_0)\left[\left(\frac{V}{V_0}\right)^{q'} -
1\right]\right\}$ & $\gamma = \gamma_0
exp\left\{\ln(2\gamma_0)\left[\left(1-\varepsilon\right)^{q'} -
1\right]\right\}$ & (\ref{Jeanloz_g1})\\[0.8ex] \\ [-1.5ex]
This work & $\gamma = \gamma_0
exp\left\{\ln\left(\frac{3}{2}\gamma_0\right)\left[\left(\frac{V}{V_0}\right)^{q'}
- 1\right]\right\}$ & $\gamma = \gamma_0
exp\left\{\ln\left(\frac{3}{2}\gamma_0\right)\left[\left(1-\varepsilon\right)^{q'}
- 1\right]\right\}$ & (\ref{Jeanloz_g2})\\[0.8ex] \\ [-1.5ex]
This work & $\gamma = \gamma_0
exp\left\{\ln\left(\frac{\gamma_0}{\gamma_{\infty}}\right)\left[\left(\frac{V}{V_0}\right)^{q'}
- 1\right]\right\}$ & $\gamma = \gamma_0
exp\left\{\ln\left(\frac{\gamma_0}{\gamma_{\infty}}\right)\left[\left(1-\varepsilon\right)^{q'}
- 1\right]\right\}$ & (\ref{my_gamma})\\[0.8ex] \\ [-1.5ex]
Rice\,\cite{Rice_1965} & $\frac{\gamma}{V}
=\frac{\gamma_0}{V_0}\left[1+\gamma_0\left(1-\frac{V}{V_0}\right)\right]^{-1}$
& $\gamma=\gamma_0(1-\varepsilon)(1+\gamma_0\varepsilon)^{-1}$ &
(\ref{Rice_1})\\

\bottomrule[1pt]
\end{tabular}
\caption{Original and transformed expressions for the Gr\"uneisen
ratio \label{all_equations}}
\end{sidewaystable}
The constant ($2\over3$) in Eqs.(\ref{LANL}) and (\ref{Sandia}) is
replaced by $\gamma_{\infty}$.

The values of the Gr\"uneisen ratio at ambient conditions
$\gamma_0$, $\gamma_{\infty}$ --- the value of $\gamma$ at $P \to
\infty$, the second and the third Gr\"uneisen ratios $q$ and $q'$,
and the material constant $a$ are the parameters to be determined
from the best fit of the experimental datasets.

\begin{figure}[!htbp]
\subfloat{\includegraphics[width=.45\linewidth]{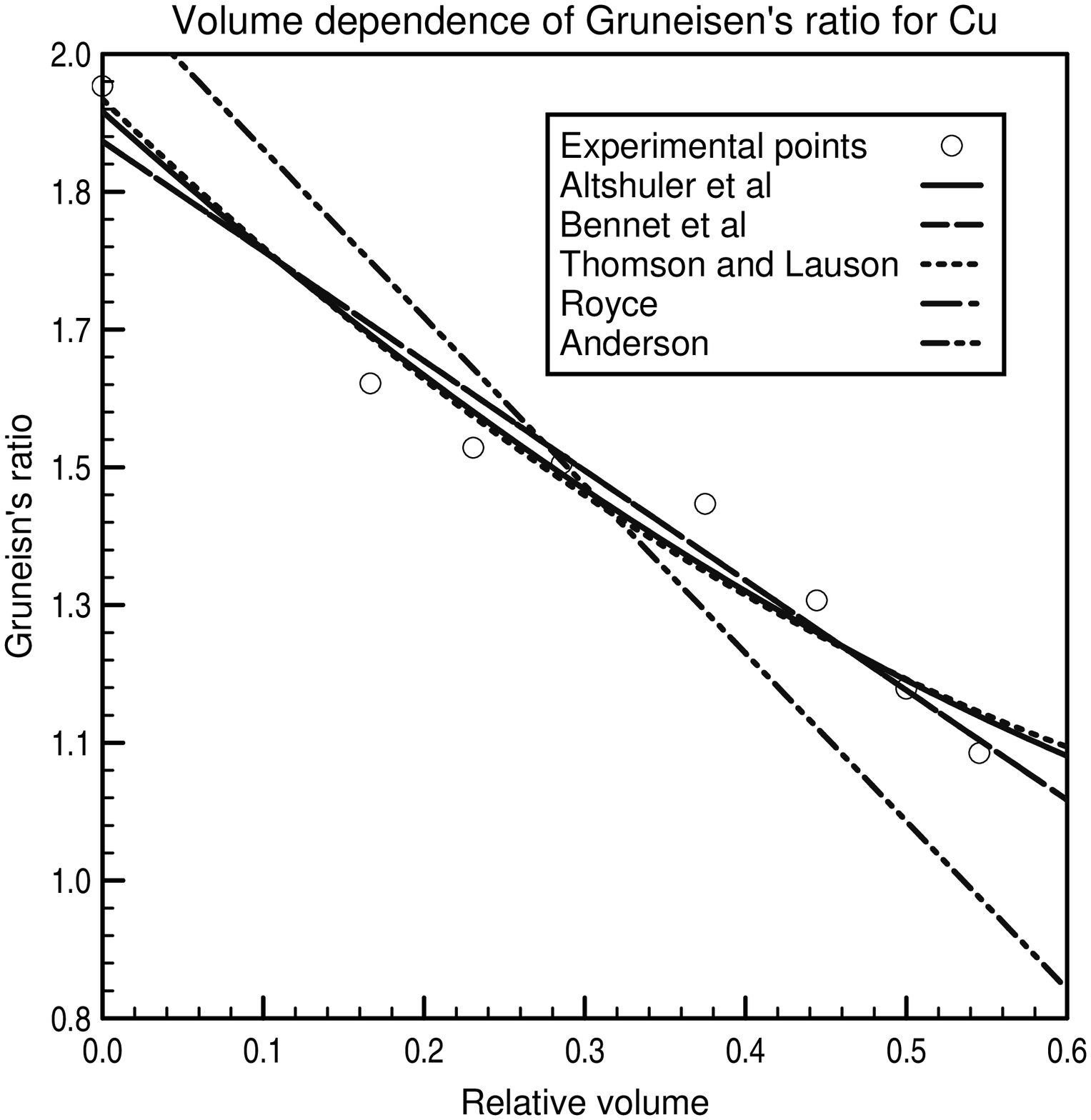}}
\subfloat{\includegraphics[width=.45\linewidth]{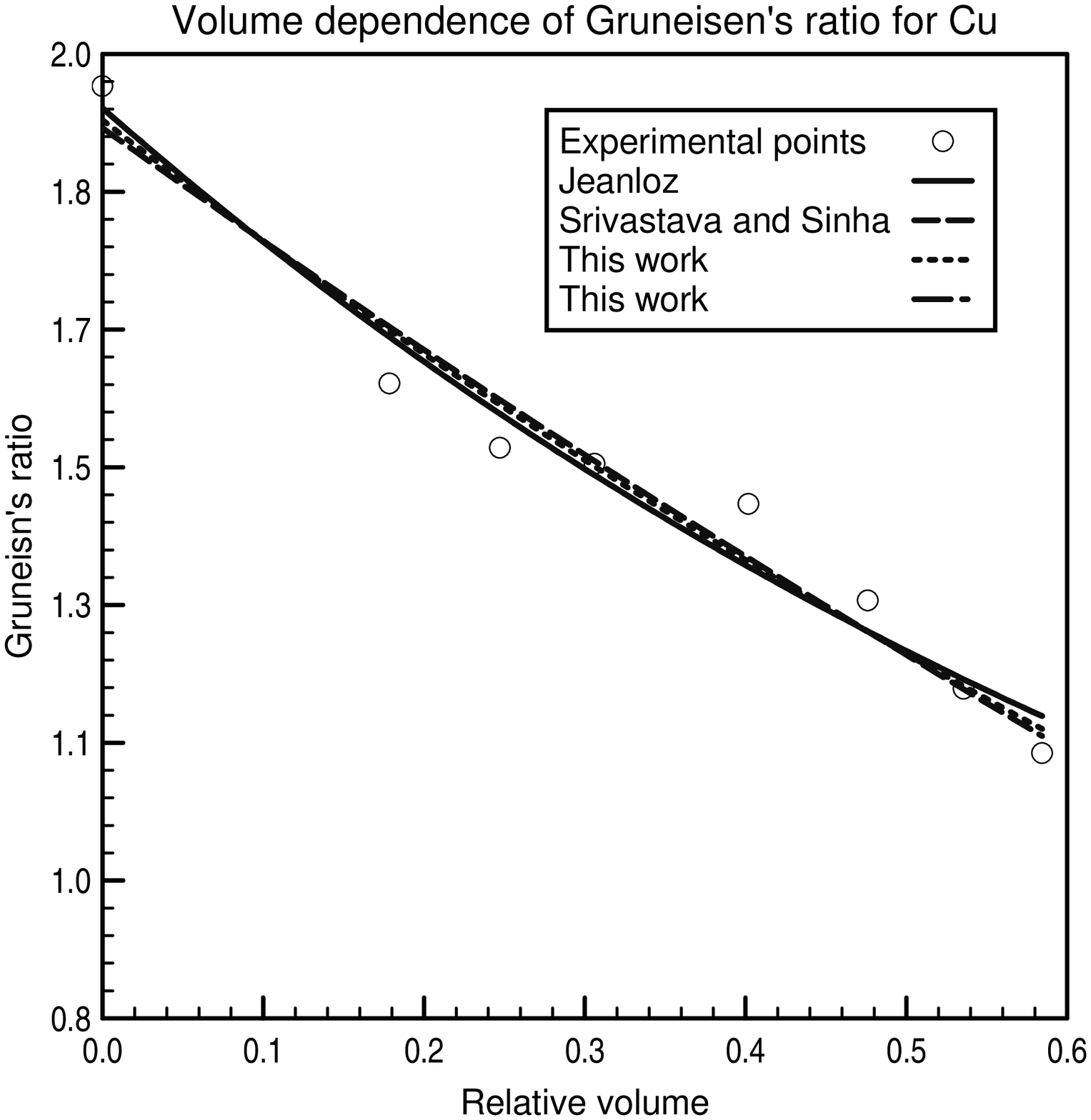}}\\
\subfloat{\includegraphics[width=.45\linewidth]{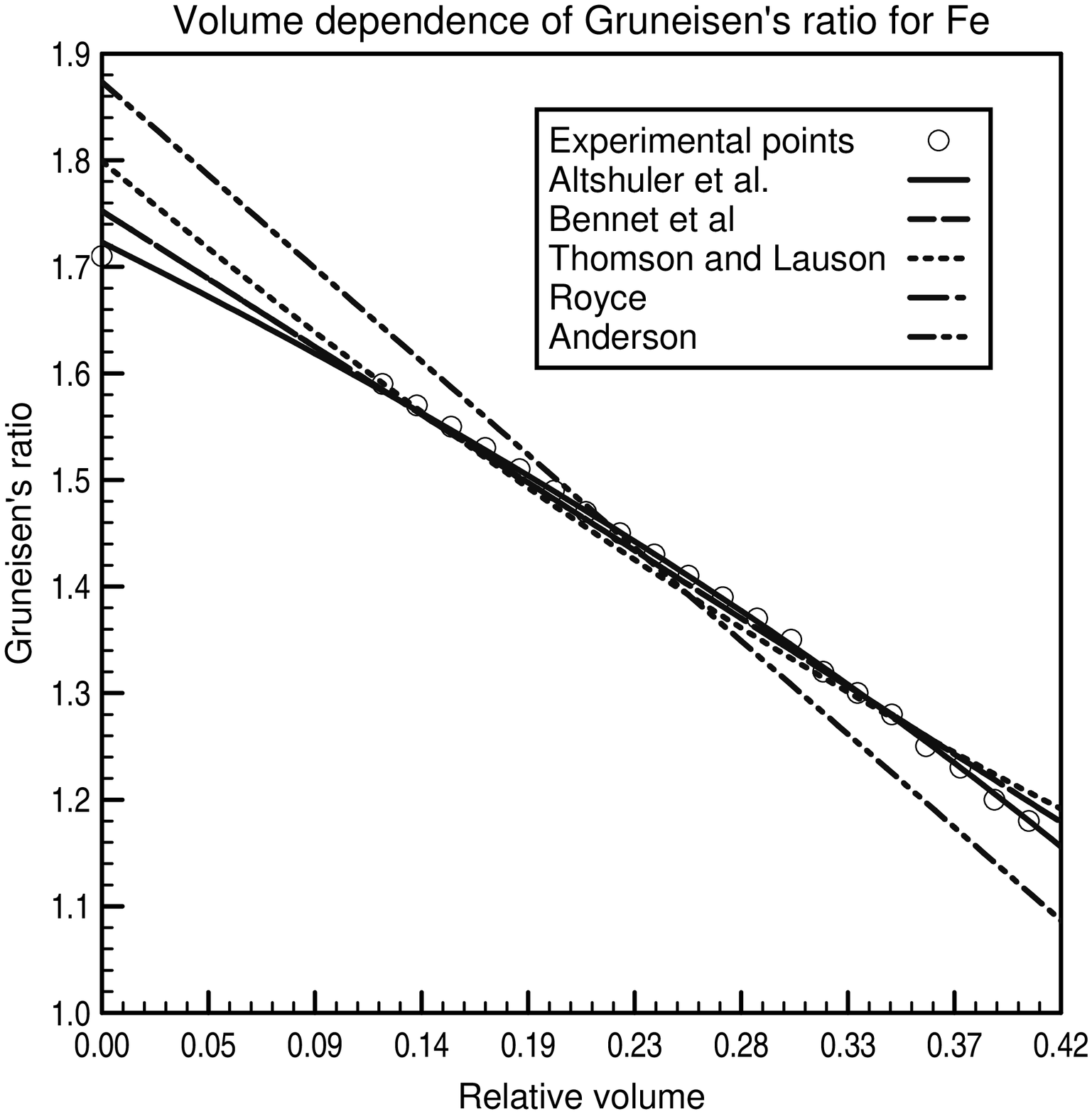}}
\subfloat{\includegraphics[width=.45\linewidth]{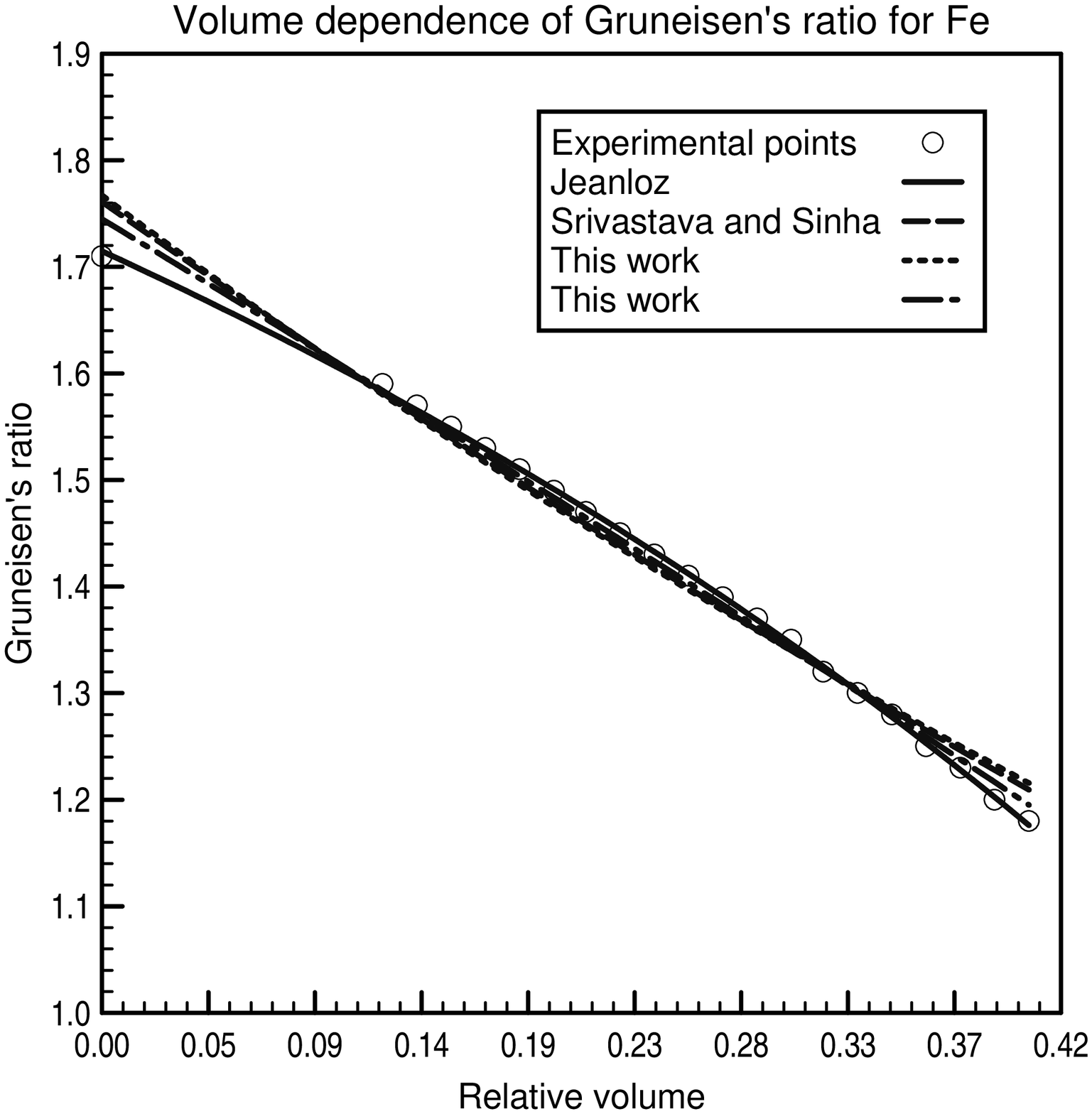}}\\
\subfloat{\includegraphics[width=.45\linewidth]{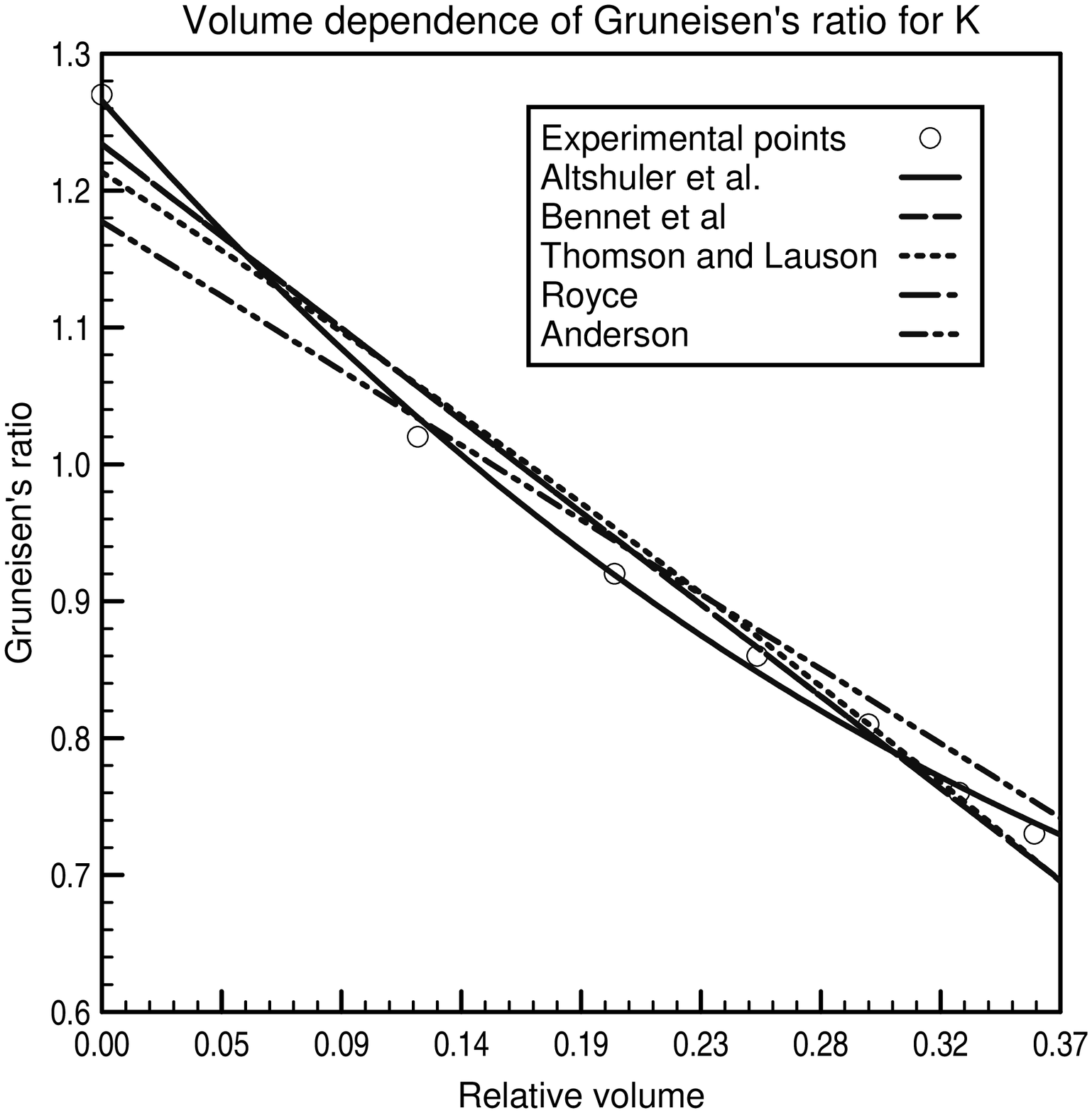}}
\subfloat{\includegraphics[width=.45\linewidth]{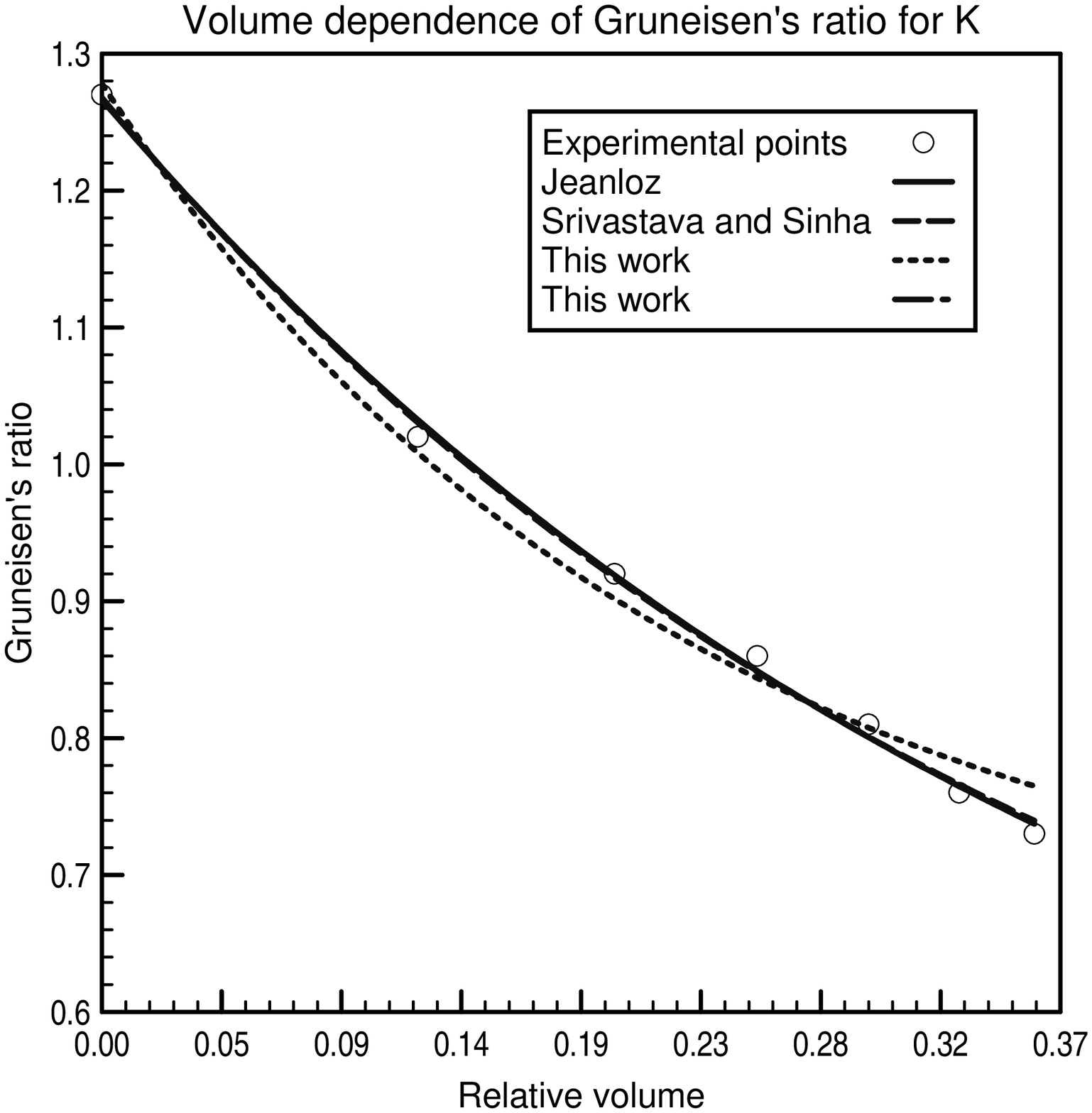}}
\caption{Volume dependence of the Gr\"uneisen ratio for $Cu$,
$\varepsilon$-$Fe$ and $K$} \label{all_metals}
\end{figure}

\begin{figure}[!htbp]
\subfloat{\includegraphics[width=.45\linewidth]{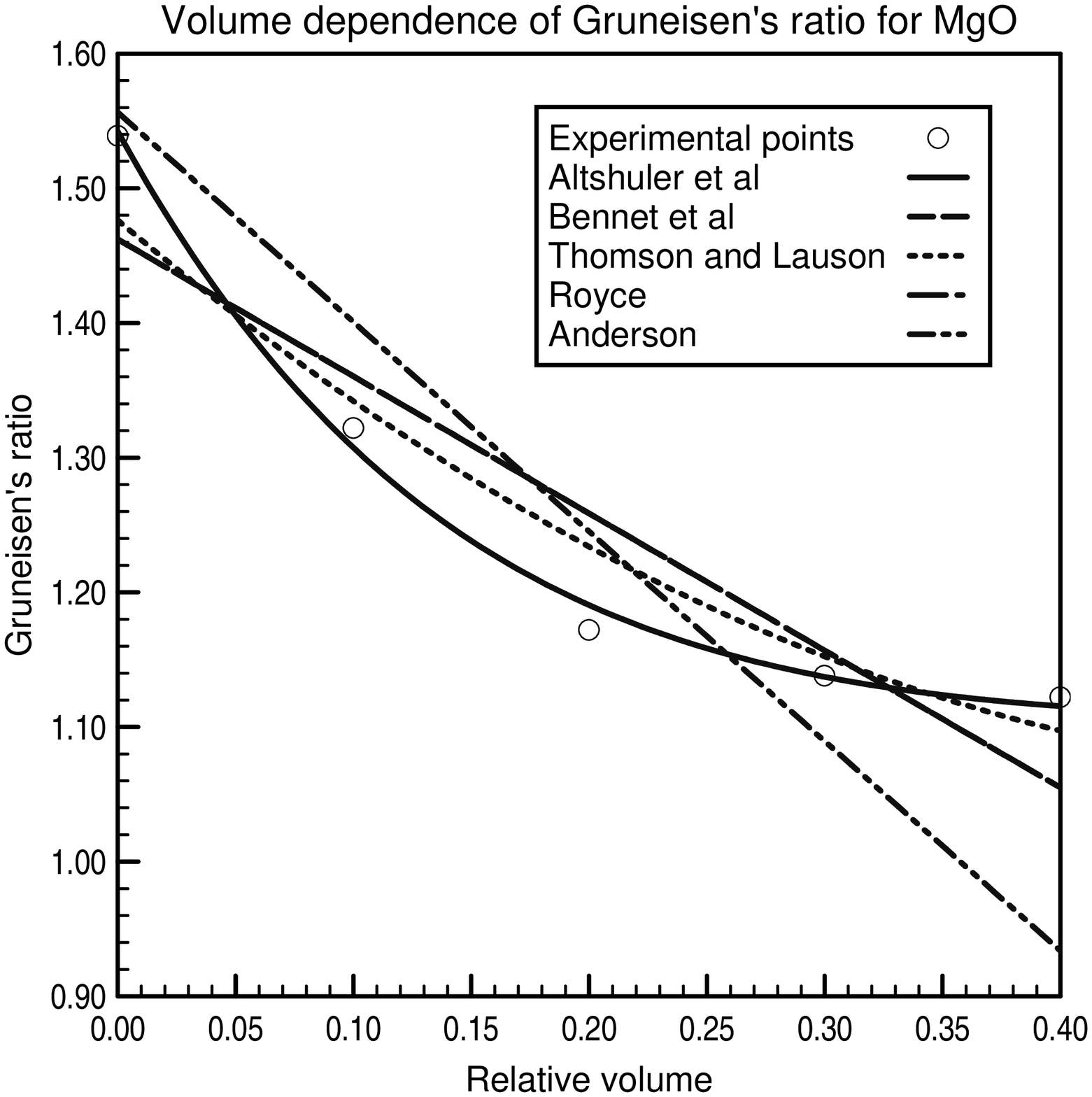}}
\subfloat{\includegraphics[width=.45\linewidth]{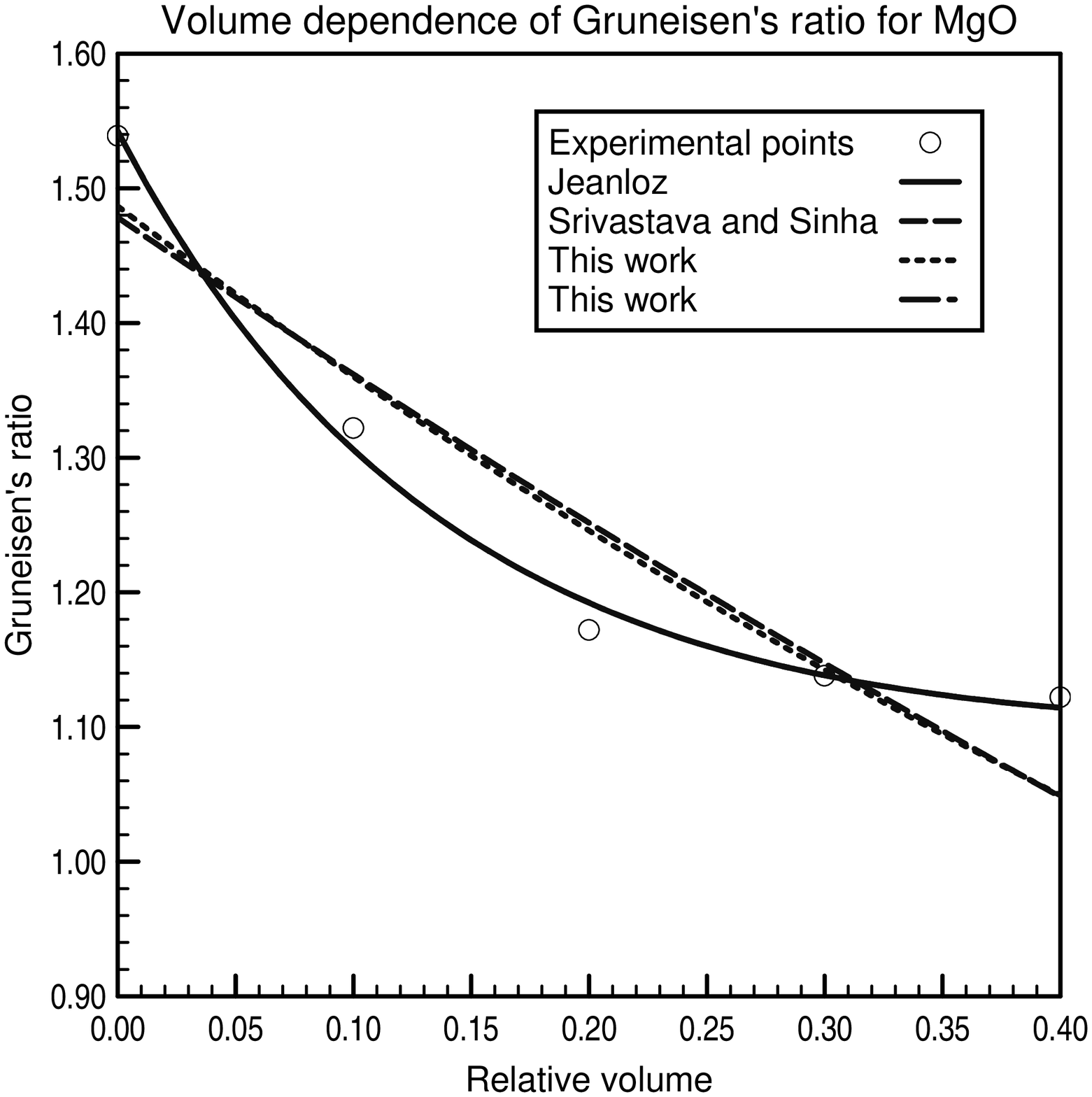}}\\
\subfloat{\includegraphics[width=.45\linewidth]{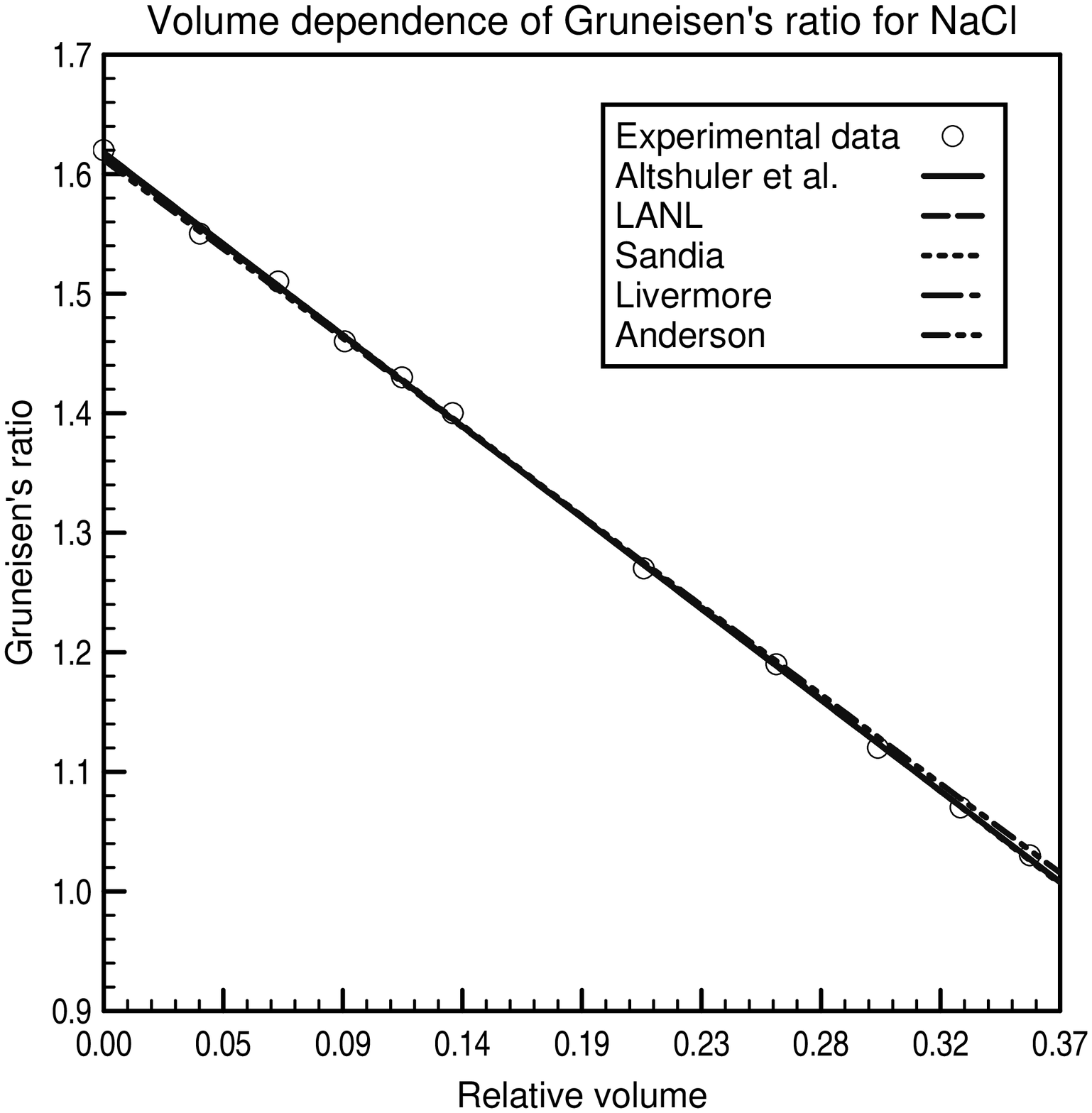}}
\subfloat{\includegraphics[width=.45\linewidth]{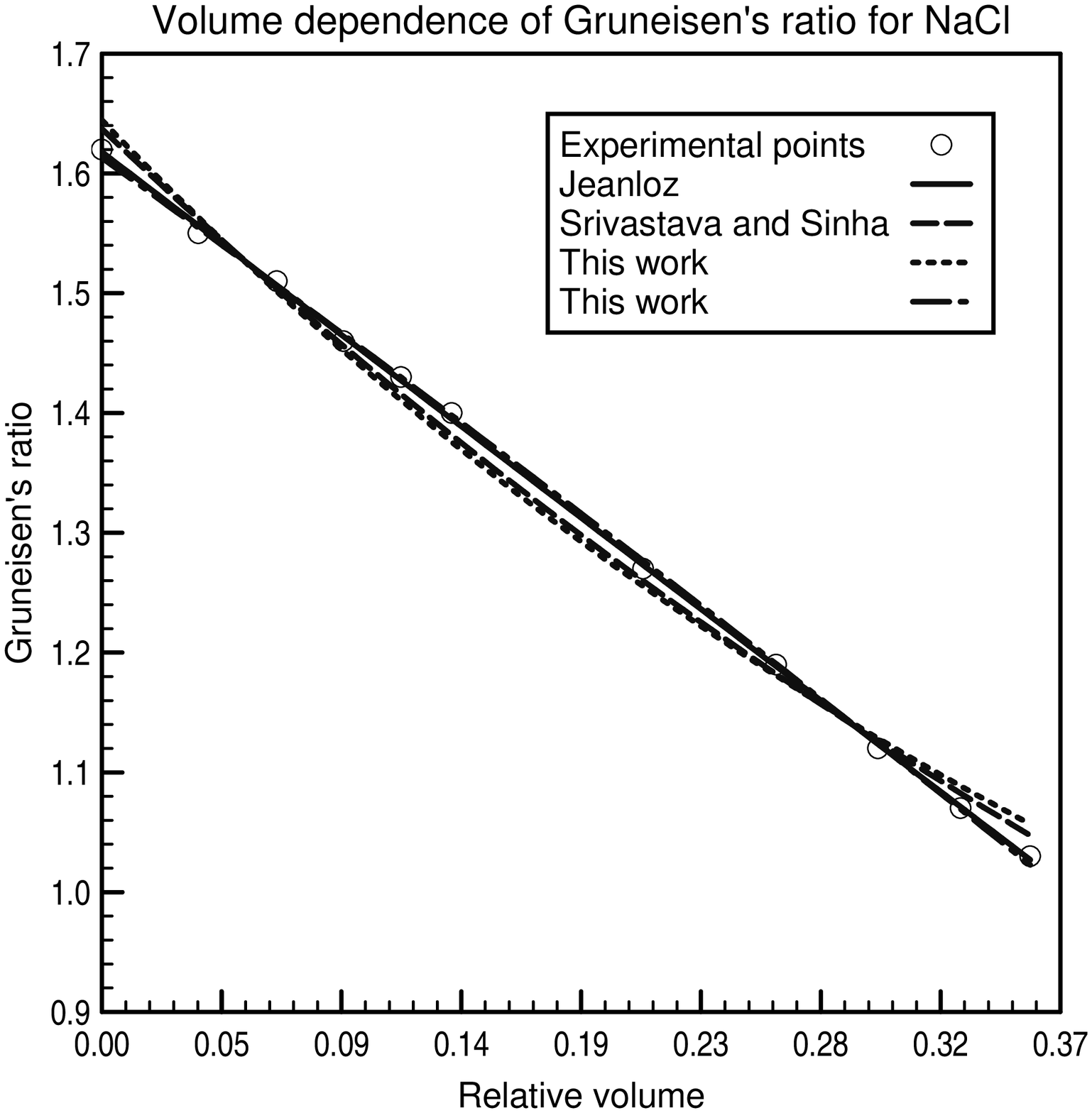}}\\

\caption{Volume dependence of the Gr\"uneisen ratio for $MgO$
 and $NaCl$} \label{all_compounds}
\end{figure}

The calculated results are presented in Tabls.(\ref{g_zero})\,-\,(4)
and in Figs(\ref{all_metals})\,-\,(\ref{all_compounds}) along with
the experimental data points for comparison.
\begin{table}
\caption{Experimental and calculated values of $\gamma_0$
\label{g_zero}}
\centering {\small
\begin{tabular}{@{}p{3cm}lllll@{}}
\toprule[1pt]
$\gamma_0$ & Cu & $\varepsilon$-Fe & K & NaCl & MgO \\
\midrule[0.6pt]
Experimental value & 2.0 & 1.71 & 1.27 & 1.62 & 1.539 \\

Anderson\,\cite{Anderson_1979} & 2.091 & 1.874 & 1.177 & 1.612 & 1.556 \\

Bennett\,{\em et al}\,\cite{Bennet_1978} & 1.891 & 1.752 & 1.234 & 1.618 & 1.462 \\

Thomson\,\,and Lauson\,\cite{Thomson_1972} & 1.944 & 1.799 & 1.213 & 1.618 & 1.476 \\

Royce\,\cite{Royce_1971} & 1.891 & 1.752 & 1.234 & 1.616 & 1.462 \\

Al'tshuler {\em et~al}\,\cite{25_metals} & 1.928 & 1.723 & 1.266 & 1.634 & 1.542 \\

Jeanloz\,\cite{Jeanloz_1989} & 1.933 & 1.715 & 1.267 & 1.618 & 1.542 \\

Srivastava\,\,and Sinha\,\cite{Srivastava} & 1.908 & 1.761 & 1.268 & 1.637 & 1.478 \\

This work & 1.918 & 1.767 & 1.278 & 1.644 & 1.487 \\

This work & 1.933 & 1.745 & 1.267 & 1.620 & 1.542 \\
\bottomrule[1pt]
\end{tabular}
}
\end{table}

From Tabls.(\ref{g_zero}) - (4) and Figs.(\ref{all_metals}) - (\ref{all_compounds}) we can see that
Eqs.(\ref{common})\,-\,(\ref{Altshuler}), (\ref{Jeanloz_g}), and
(\ref{Jeanloz_g1})\,-\,(\ref{my_gamma}) are in good agreement with
the experimental datasets. In all cases Eqs.(\ref{Altshuler}),
(\ref{Jeanloz_g}), and (\ref{my_gamma}) have the highest and
practically coinciding coefficients of multiple determination $R^2$
and the smallest error in $\gamma$. The errors in $\gamma$ for the
other expressions are within the range of the experimental errors
and the coefficients of multiple determination $R^2$ are high enough
for the models to be considered adequate.
\begin{table}[!htb]
\caption{Coefficient of multiple determination $R^2$ and error in
$\gamma$ [\%] for {\it Cu}, {\it $\varepsilon$-Fe}, and {\it K}
\label{metals} }
\centering {\small
\begin{tabular}{@{}p{2.5cm}llllll@{}}
\toprule[1pt] \multicolumn{1}{l}{\multirow{2}{*}{Equations}} &
\multicolumn{2}{c}{Cu} & \multicolumn{2}{c}{$\varepsilon$-Fe} &

\multicolumn{2}{c}{K} \\
\cmidrule[0.6pt](lr){2-7}
 & \multicolumn{1}{c}{$R^2$} & \multicolumn{1}{c}{Error in $\gamma_0$ [\%]} & \multicolumn{1}{c}{$R^2$} &

\multicolumn{1}{c}{Error in $\gamma_0$ [\%]} & \multicolumn{1}{c}{$R^2$} & \multicolumn{1}{c}{Error in $\gamma_0$ [\%]} \\
\midrule[1pt] Al'tshuler {\em et~al}\,\cite{25_metals} &
\multicolumn{1}{c}{0.964} & \multicolumn{1}{c}{1.611} &
\multicolumn{1}{c}{0.999}

& \multicolumn{1}{c}{0.766} & \multicolumn{1}{c}{0.997} & \multicolumn{1}{c}{0.336} \\

Bennett\,{\em et al}\,\cite{Bennet_1978} & \multicolumn{1}{c}{0.957}
& \multicolumn{1}{c}{3.505} & \multicolumn{1}{c}{0.991}

& \multicolumn{1}{c}{2.482} & \multicolumn{1}{c}{0.982} & \multicolumn{1}{c}{2.842} \\

Thomson\,\,and Lauson\,\cite{Thomson_1972} &
\multicolumn{1}{c}{0.963} & \multicolumn{1}{c}{0.828} &

\multicolumn{1}{c}{0.969} & \multicolumn{1}{c}{5.226} & \multicolumn{1}{c}{0.969} & \multicolumn{1}{c}{4.453} \\

Royce\,\cite{Royce_1971} & \multicolumn{1}{c}{0.957} &
\multicolumn{1}{c}{3.505} & \multicolumn{1}{c}{0.991} &

\multicolumn{1}{c}{2.482} & \multicolumn{1}{c}{0.982} & \multicolumn{1}{c}{2.842} \\

Anderson\,\cite{Anderson_1979} & \multicolumn{1}{c}{0.670} &
\multicolumn{1}{c}{6.667} & \multicolumn{1}{c}{0.852} &

\multicolumn{1}{c}{9.562} & \multicolumn{1}{c}{0.945} & \multicolumn{1}{c}{7.299} \\

Jeanloz\,\cite{Jeanloz_1989} & \multicolumn{1}{c}{0.966} &
\multicolumn{1}{c}{5.634} & \multicolumn{1}{c}{0.999} &

\multicolumn{1}{c}{0.342} & \multicolumn{1}{c}{0.998} & \multicolumn{1}{c}{1.067} \\

Srivastava\,\,and Sinha\,\cite{Srivastava} &
\multicolumn{1}{c}{0.962} & \multicolumn{1}{c}{5.437} &

\multicolumn{1}{c}{0.984} & \multicolumn{1}{c}{1.828} & \multicolumn{1}{c}{0.998} & \multicolumn{1}{c}{0.972} \\

This work & \multicolumn{1}{c}{0.964} & \multicolumn{1}{c}{5.263} &
\multicolumn{1}{c}{0.979} & \multicolumn{1}{c}{2.126} &

\multicolumn{1}{c}{0.988} & \multicolumn{1}{c}{2.259} \\

This work & \multicolumn{1}{c}{0.966} & \multicolumn{1}{c}{5.634} &
\multicolumn{1}{c}{0.994} & \multicolumn{1}{c}{1.162} &

\multicolumn{1}{c}{0.998} & \multicolumn{1}{c}{1.067} \\
\bottomrule[1pt]
\end{tabular}
}
\end{table}

Equation\,(\ref{common}) stands aside in this classification. It has
noticeably lower $R^2$ and is the worst in all cases except $NaCl$.
The volume variation of $\gamma$ for $NaCl$ is adequately described
by all models, although a slight departure of the '($\gamma/V) =
const$' approximation can be observed as $\varepsilon$ increases
(Fig.(\ref{all_compounds})).

It can be readily seen from Figs.(\ref{all_metals}) and
(\ref{all_compounds}) that in all cases Eq.(\ref{LANL}) and
Eq.(\ref{LLNL}) completely overlap. This might be explained by the
fact that both expressions are represented by a linear model.

\begin{table}[!htb]
\caption{Coefficient of multiple determination $R^2$ and error in
$\gamma$ [\%] for {\it NaCl} and {\it MgO} \label{compounds}}
\centering {\small
\begin{tabular}{@{}p{3cm}llll@{}}
\toprule[1pt]
\multicolumn{1}{l}{\multirow{2}{*}{Equations}} & \multicolumn{2}{c}{NaCl} & \multicolumn{2}{c}{MgO} \\
\cmidrule[0.6pt](lr){2-5}
 & $R^2$ & Error in $\gamma$ [\%] & $R^2$ & Error in $\gamma$ [\%] \\
\midrule[1pt]

Al'tshuler {\em et~al}\,\cite{25_metals} & \multicolumn{1}{c}{0.999}
& \multicolumn{1}{c}{0.12\phantom{0}} &

\multicolumn{1}{c}{0.995} & \multicolumn{1}{c}{0.21\phantom{0}} \\
 Bennett\,{\em et al}\,\cite{Bennet_1978} &
\multicolumn{1}{c}{0.999} & \multicolumn{1}{c}{0.14\phantom{0}} &

\multicolumn{1}{c}{0.84\phantom{0}} & \multicolumn{1}{c}{4.99\phantom{0}} \\
 Thomson\,\,and Lauson\,\cite{Thomson_1972} &
\multicolumn{1}{c}{0.999} & \multicolumn{1}{c}{0.26\phantom{0}} &

\multicolumn{1}{c}{0.927} & \multicolumn{1}{c}{4.07\phantom{0}} \\
 Royce\,\cite{Royce_1971} & \multicolumn{1}{c}{0.999} &
\multicolumn{1}{c}{0.14\phantom{0}} &

\multicolumn{1}{c}{0.84\phantom{0}} & \multicolumn{1}{c}{4.99\phantom{0}} \\
 Anderson\,\cite{Anderson_1979} & \multicolumn{1}{c}{0.999} &
\multicolumn{1}{c}{0.5\phantom{00}} & \multicolumn{1}{c}{0.598}

& \multicolumn{1}{c}{1.13\phantom{0}} \\
 Jeanloz\,\cite{Jeanloz_1989} & \multicolumn{1}{c}{0.999} &
\multicolumn{1}{c}{0.423} & \multicolumn{1}{c}{0.994} &

\multicolumn{1}{c}{1.922} \\
 Srivastava\,\,and Sinha\,\cite{Srivastava} &
\multicolumn{1}{c}{0.996} & \multicolumn{1}{c}{1.378} &

\multicolumn{1}{c}{0.861} & \multicolumn{1}{c}{7.562} \\

This work & \multicolumn{1}{c}{0.992} & \multicolumn{1}{c}{1.916} & \multicolumn{1}{c}{0.879} & \multicolumn{1}{c}{7.059} \\

This work & \multicolumn{1}{c}{0.999} & \multicolumn{1}{c}{0.438} & \multicolumn{1}{c}{0.994} & \multicolumn{1}{c}{1.922} \\
\bottomrule[1pt]
\end{tabular}
}
\end{table}

{\sl{The infinite compression limit of $\gamma$.}} Four of the
considered  models --- Eqs.(\ref{LANL}), (\ref{Sandia}),
(\ref{Altshuler}), and (\ref{Jeanloz_g})  contain $\gamma_{\infty}$
(the value of $\gamma$ at P$\rightarrow\infty$). It is assumed that
at infinite pressure ($P\rightarrow\infty$)  solids become a
crystalline one-component plasma, i.e. an oscillating lattice of
ions in a uniform neutralizing background of electrons
\cite[Ch.~17]{Young_Ph_Dgrms}. A number of theoretical works predict
$\gamma$ = $1\over2$ for this limiting state of a solid. Kopyshev
\cite{Kopyshev_1965} calculated $\gamma(V)$ in the Thomas-Fermi
approximation and found $\gamma$ = $1\over2$ as
$P\rightarrow\infty$. Various theoretical studies by other authors
\cite{Holt_1970, Ross_1985, Nagara_1985} as well as simple
dimensional arguments by Hubbard \cite[p.~34]{Hubbard_1984} also
lead to $\gamma$ = $1\over2$ as $P\rightarrow\infty$. Other
researchers\,\cite[and references cited therein]{Holzapfel_2001}
consider ($2\over3$) a more appropriate value of $\gamma_{\infty}$
for solids due to the fact that the linear temperature dependence of
the electronic specific heat of the degenerate free electron gas
dominates over the phonon contribution when the Debye temperature is
increased sufficiently. Al'tshuler {\sl et al}\,\cite{25_metals}
assume $2\over3$ to be the infinite compression limit of $\gamma$
for all materials except alkali metals, for which
$\gamma_{\infty}$=$1\over2$.

Unfortunately, none of the expressions for $\gamma(V)$, considered
in the present work follow either of these constraints at infinite
pressure. This can be readily seen from Tabl.(\ref{g_inf}), where
there are given the values of $\gamma_{\infty}$, obtained by
regression analysis.
There are several exceptions. Equation\,(\ref{LANL}), proposed by
Bennet {\sl et al}\,\cite{Bennet_1978} predicts for $Cu$ and $MgO$
values of $\gamma_{\infty}$ near $1\over2$.
Equation\,(\ref{Altshuler}), proposed by Al'tshuler {\sl et
al}\,\cite{25_metals} and Eq.(\ref{my_gamma}) (this work) predict
for $K$ values of $\gamma_{\infty}$  close to $1\over2$.
\begin{table}[!htb]
\caption{Calculated values of $\gamma_{\infty}$ \label{g_inf}}
\centering {\small
\begin{tabular}{@{}p{3cm}lllll@{}}
\toprule[1pt]
$\gamma_{\infty}$ & Cu & $\varepsilon$-Fe & K & NaCl & MgO \\
\midrule[0.6pt]
Bennett\,{\em et al}\,\cite{Bennet_1978} & 0.525 & \phantom{-}0.387 & -0.219 & -3.119 & 0.444 \\

Thomson\,\,and Lauson\,\cite{Thomson_1972} & 1.001 & \phantom{-}0.838 & -0.504 & -0.085 & 1.321 \\

Al'tshuler {\em et~al}\,\cite{25_metals} & 0.936 & -8.423 & \phantom{-}0.552 & -1.511 & 1.105 \\

This work & 0.858 & \phantom{-}2.0$\times$10$^{-4}$ & \phantom{-}0.474 & \phantom{-}3.1$\times$10$^{-7}$ & 1.1 \\
\bottomrule[1pt]
\end{tabular}
}
\end{table}

Srivastava and Sinha in their paper\,\cite{Srivastava} have noticed
that an infinite compression limit may be obtained from
Eq.(\ref{Jeanloz_g}), i.e. Eq.(\ref{g_infty}). The values of
$\gamma_{\infty}$, calculated from Eq.(\ref{g_infty}) are also far
from theoretical predictions.

\section{Conclusions}
\label{conclusions}

The most frequently used self-contained expressions for the volume
dependence of the Gr\"uneisen ratio have been considered in the
present work and compared to available experimental data for $Cu$,
$\varepsilon$-$Fe$, $K$, $MgO$, and $NaCl$.

All expressions predict with very good accuracy values for $\gamma$
at ambient conditions. The '$\gamma\rho = const$' approximation
fails at higher compressions. The expressions proposed by Bennet
{\sl et al} \cite{Bennet_1978} and by Royce \cite{Royce_1971} are
equivalent. The model proposed by Jeanloz \cite{Jeanloz_1989}, its
modification in the present work, and the expression of Al'tshuler
{\sl et al} \cite{25_metals} are the best fits to the experimental
data sets. In the author's opinion preference should be given to
Jeanloz's model (and its modification, proposed here) because it is
based on physical assumptions, whilst the expression of  Al'tshuler
{\sl et al} is just an interpolation formula. In view of its
possible application to deriving a complete EOS for solids from
their response to shock-wave loading Eq.(\ref{my_gamma}) is more
convenient to use than Eq.(\ref{Jeanloz_g}) because it contains
$\gamma_{\infty}$ instead of $q_0$, which is not frequently used in
shock physics.

With a few exceptions, none of the models, considered here, predict
correct values for $\gamma_{\infty}$. According to Young
\cite[Ch.\,17]{Young_Ph_Dgrms} matter approaches its infinite compression
state when $\rho/\rho_0\sim$10, or, in terms of relative volume
$\varepsilon\sim$0.9. If we accept this criterion, we could say that
the experimental data sets, used in this work are nearer to
the origin of the pressure axis than to $P\to\infty $. In the case
of $\varepsilon$-$Fe$\,\cite{Anderson_2001} at $P=359.5\,GPa$ (the
highest pressure in the experiments considered here)
$\rho/\rho_0=1.684$. That is why the predictions for
$\gamma_{\infty}$ from the regression analysis are not good. In the
author's opinion the values for $Cu$ and $MgO$, obtained from the
expression of Bennet {\sl et al} (Eq.(\ref{LANL})) are sooner random
results than correctly predicted values. The case of $K$ is somewhat
different. It has a very small bulk modulus --- 3.09\,$GPa$
\cite{Boehler_1983}, i.e. very large compressibility. It is
reasonable to accept that the applied pressure drives it nearer to
the maximum compression state than the same pressure, applied to the
other materials considered here. Also, the values near ($1\over2$)
are predicted by two of the equations, having the highest
coefficient of multiple determination.

It is obvious that experiments at higher pressures are necessary to
determine more reliably the infinite compression limit of $\gamma$.
Computer simulations easily surmount the limitations of laboratory experiments.
They could be used to clarify the ability of the considered models to predict
the infinite compression limit of $\gamma$.

These inferences trace out a possible line for continuation of the
present research. A regression analysis of results from computer
simulations, using the models, considered here, might give more
reliable values of  $\gamma_{\infty}$.

This work was partially presented at the Seventh scientific
conference with international participation “SPACE, ECOLOGY,
SAFETY”, 29 November – 1 December 2011, Sofia, Bulgaria.


\begin{thebibliography}{99}
\setlength{\itemsep}{-1.5mm}
\bibitem{Girifalco_2000} Girifalco, L.A. {\it Statistical mechanics of solids}, Oxford University Press: New York, 2000.
\bibitem{Anderson_EoS} Anderson, O. L. {\it Equations of state of solids for geophysics and ceramic science}, Oxford University Press: New York, 1995.
\bibitem{Vocadlo_2000} Vocadlo L., J.P. Poirer, and G.D. Price, "Gr\"uneisen parameters and isothermal equations of state" {\em Am. Mineral}., 85(2), 390–395 (2000).
\bibitem{Knopoff_rev} Knopoff L. and J. N. Shapiro, "Comments on the Interrelationships between Gr\"uneisen's Parameter and Shock and Isothermal Equations of State", {\em J.Geophys.Res.}, 74, 1439-1450 (1969).
\bibitem{Anderson_rev} Anderson O. L., "The Gr\"uneisen ratio for the last 30 years", {\em Geophys. J. Int.}, 143, 279-294 (2000).
\bibitem{Peng_2007} Peng X., L. Xing, and Zh. Fang, "Comparing research on the pressure or volume dependence of Gr\"uneisen parameter", {\em Physica B}, 394, 111-114 (2007).
\bibitem{Guang-lei} Guang-lei Cui and Rui-lan Yu, "Volume and pressure dependence of the Gr\"uneisen parameter $\gamma$ for solids at high temperature", {\em Physica B}, 390, 220-224 (2007).
\bibitem{Anderson_1979} Anderson O. L., "Evidence Supporting the Approximation $\gamma\rho=const$ for the Gr\"uneisen parameter of the Earth's Lower Mantle", {\em J.Geophys.Res.}, 84, 3537-3542 (1979).
\bibitem{Bennet_1978} Bennet B.I., J.D. Johnson, G.I. Kerley, and G.T. Root, {\it Los Alamos Scientific Laboratory report LA--7130} (1978).
\bibitem{Thomson_1972} Thomson S.L. and H.S. Lauson, {\it Sandia Livermore Corporation report SC-RR-710714} (1972).
\bibitem{Royce_1971} Royce E. {\it Lawrence Livermore Laboratory report UCRL 51121} (1971).
\bibitem{Holzapfel_2001} Holzapfel, W. B., Hartwig, M., and Sievers, W. “Equations of state for Cu, Ag, and Au for wide ranges in temperature and pressure up to 500 GPa and above”, {\it J. Phys. Chem. Ref. Data}, {\bf 30}\,(2), 515-530, (2001).
\bibitem{25_metals} Altshuler L.V., S.E. Brusnikin, and E.A. Kuzmenkov, "Isotherms and Gr\"uneisen Functions for 25 Metals", {\em Zhurnal Prikladnoi Mekhaniki i Tekhnicheskoi Fiziki}, 1987, No. 1, pp. 134-146. (English trans., {\em Journal of Applied Mechanics and Technical Physics}, Vol. 28, No. 1, 1987, pp. 129-141)
\bibitem{Jeanloz_1989} Jeanloz, R., "Shock wave equation of state and finite strain theory", {\em J.Geophys.Res.}, 94(B5), 5873-5886 (1989).
\bibitem{Srivastava} Srivastava S.K. and P. Sinha, "Analysis of volume dependence of Gr\"uneisen ratio", {\em Physica B}, 404, 4316-4320 (2009).
\bibitem{Rice_1965} Rice, M.H., “Pressure-volume relations for the alkali metals from shock-wave measurements”, {\em J. Phys. Chem. Solids}, 26, 483-492, (1965).
\bibitem{Birch_1986} Birch F., "Equation of State and Thermodynamic Parameters of NaCl to 300 kbar in the High-Temperature Domain", {\em J.Geophys.Res.}, 91, 4949-4954 (1986).
\bibitem{Slater_1939bk} Slater, J.C. {\it Introduction to Chemical Physics}, McGraw-Hill: New York, 1939; Ch.XIV, p.239.
\bibitem{Landau_1945} Landau, L. D., and K. P. Stanyukovich. "Ob izuchenii detonacii kondensirovannyx vzryvchatyx veshchestv",  {\it Dokl. AN SSSR}, {\bf 46}, 112-117, (1945). [in Russian]
\bibitem{Slater_1940} Slater J.C., “Note on Gr\"uneisen's Constant for the Incompressible Metals”, {\it Phys.Rev.} 57,  744-746 (1940).
\bibitem{Gilvarry_1956} Gilvarry J.J., “Gr\"uneisen Parameter for a Solid under Finite Strain”, {\it Phys.Rev.} 102(2),  331-340 (1956).
\bibitem{Dugdale_1953} Dugdale J. S. and D. K. C. Macdonald, “The Thermal Expansion of Solids”, {\it Phys.Rev.} 89(4), 832-834 (1953).
\bibitem{Rice_1958} Rice M. H., R. G. McQueen, and J. M. Walsh , “Compression of Solids by Strong Shock Waves“, {\it Sol.St.Phys.} 6, 1-63 (1958).
\bibitem{Vaschenko_1963} Vaschenko V.Y. and V.N. Zubarev, “Concerning the Gr\"uneisen constant”, {\it Sov. Phys. Sol. St.} 5, 653-655 (1963).
\bibitem{Anderson_2001} Anderson, O. L., L. Dubrovinsky, S. K. Saxena, and T. LeBihan, "Experimental vibrational Gr\"uneisen ratio values for $\varepsilon$-iron up to 330 GPa at 300 K", {\em Geophys. Res. Lett.}, 28(2), 399-402 (2001).
\bibitem{Anderson_2001c} Anderson, O. L., L. Dubrovinsky, S. K. Saxena, and T. LeBihan, Correction to “Experimental vibrational Gr\"uneisen ratio values for $\varepsilon$-iron up to 330 GPa at 300~K”, {\em Geophys. Res. Lett.}, 28(12), 2359 (2001).
\bibitem{Boehler_1983} Boehler R., "Melting temperature, adiabats, and Gr\"uneisen parameter of lithium, sodium and potassium versus pressure", {\em Phys. Rev. B} 27, 6754 (1983).
\bibitem{Anderson_1993} Anderson O. L., H. Oda, A. Chopelas, and D. G. Isaak, "A thermodynamic theory of the Gr\"uneisen ratio at extreme conditions: MgO as an example", {\em Phys.Chem.Min}, 19(6), 369-380 (1993).
\bibitem{Young_Ph_Dgrms} Young, D. A. {\it Phase diagrams of the elements}. University of California Press: Berkeley, 1991.
\bibitem{Kopyshev_1965} Kopyshev, V. P. "Konstanta Gryunajzena v priblizhenii Tomasa – Fermi",  {\it Dokl. AN SSSR},  {\bf 161}(5), 1067-1068, (1965). [in Russian]
\bibitem{Holt_1970} Holt, A. C., and Ross, M. “Calculations of the Gr\"uneisen parameter for some models of the solid”,  {\it Phys. Rev. B}, {\bf 1}(6), 2700, (1970).
\bibitem{Ross_1985} Ross, M. “Matter under extreme conditions of temperature and pressure”, {\it Rep. Progr. Phys.}, {\bf 48}\,(1), 1, (1985).
\bibitem{Nagara_1985} Nagara, H., and Nakamura, T. “Theory of lattice-dynamical properties of compressed solids”, {\it Phys. Rev. B}, {\bf 31}\,(4), 1844, (1985).
\bibitem{Hubbard_1984} Hubbard, W. B. {\it Planetary interiors}. Van Nostrand Reinhold Co: New York, 1984, 343~p.


\end{thebibliography}
\end{document}